\documentclass[preprint,showpacs,nofootinbib,prd,aps]{revtex4-1}

\usepackage{graphicx}
\usepackage{amstext}
\usepackage{amssymb}
\usepackage[usenames]{color}

\begin{document}
\title{Particle production at energies available at the CERN Large Hadron Collider within evolutionary model}

\author{Yu.~M.~Sinyukov$^{1}$}
\author{V.~M.~Shapoval$^1$} 
\affiliation{$^1$Bogolyubov Institute for Theoretical Physics,
Metrolohichna  14b,  Kiev 03680  Ukraine
}

\begin{abstract}
The particle yields and particle number ratios in Pb+Pb collisions at the LHC energy $\sqrt{s_{NN}}=2.76$~TeV are 
described within the integrated hydrokinetic model (iHKM) at the two different equations of state (EoS) for the 
quark-gluon matter and the two corresponding hadronization temperatures, $T=165$~MeV and $T=156$~MeV. 
The role of particle interactions at the final afterburner stage of the collision in the particle production
is investigated by means of comparison of the results of full iHKM simulations with those where the annihilation
and other inelastic processes (except for resonance decays) are switched off after hadronization/particlization, 
similarly as in the thermal models. An analysis supports the picture of continuous chemical freeze-out in the 
sense that the corrections to the sudden chemical freeze-out results, which arise because of the inelastic 
reactions at the subsequent evolution times, are noticeable and improve the description of particle %yields and 
number ratios. An important observation is that although the particle number ratios with switched-off inelastic 
reactions are quite different at different particlization temperatures which are adopted for different equations of state to reproduce experimental data, the complete iHKM calculations bring very close results in both cases. 
\end{abstract}

\pacs{13.85.Hd, 25.75.Gz}
\maketitle

Keywords: {\small \textit{lead-lead collisions, LHC, particle yield, particle number ratio, freeze-out}}

%Corresponding author: {\small \textit{Yu.M. Sinyukov, Bogolyubov Institute
%for Theoretical Physics, Kiev 03680, Metrolohichna 14b, Ukraine. E-mail:
%sinyukov@bitp.kiev.ua}}

\section{Introduction}

The analysis of the particle number ratios is carried out successfully in thermal models for different energies of 
$A+A$ collisions, from the AGS to the LHC energies~\cite{pbm1, pbm2, pbm3,Becat, Clemans,pbm4, pbm5}. The thermal 
models suppose that at some hypersurface characterized by uniform temperature and baryon chemical 
potential, the chemical composition of the hadron matter is frozen out, and in subsequent evolution of the hadron 
matter the particle yield is changed only because of the resonance decays. At the LHC energies the afterburner 
``post-freeze-out'' stage  is the longest, and so there is a special interest to check the chemical freeze-out 
hypotheses within the dynamical models for these energies. The ALICE Collaboration has already 
noted~\cite{ALICE1, ALICE2} that annihilation processes at the afterburner stage, which are taken into account in 
HKM model~\cite{Uniform} noticeably improve agreement with (anti)proton spectra/yield at the LHC. The analysis of the role of inelastic processes at post-hydrodynamic stage in formation of the particle yield is continuous (see, e.g., \cite{B}).

It seems that continuous chemical freeze-out as well as kinetic freeze-out is an inevitable feature of the 
dynamical models of $A+A$ collisions since sudden chemical freeze-out means instant transition from extremely fast 
chemically equilibrated expansion (presupposing a very intensive inelastic reactions) to the evolution with 
totally forbidden inelastic reactions. 
Sudden kinetic freeze-out means an instant change of hadron cross-section from a very large one (typical for near 
perfect hydrodynamics) to zero cross-section (free streaming particles). 
Such sudden transitions are not typical for realistic dynamical models \footnote{Note also that neither first 
order phase transition, nor crossover are sudden in time in the process of system expansion.}. In our very recent 
note \cite{KstarOur} we found, using $K^{*}(892)$ probe, that at the LHC energies a good agreement with the 
experimental data for these resonances requires a relatively long kinetic freeze-out, near 5~fm/$c$ 
after particlization/hadronization. This is worth noting that continuous thermal freeze-out means not only successive freeze-out for different hadrons (as in, e.g., Ref. \cite{C}), but continuous particle emission for each species, see new important details in Ref. \cite{KstarOur}.   

In this study we calculate the particle %yields and 
number ratios in the integrated hydrokinetic model (iHKM) and compare the results with the ones obtained in thermal models. Also we calculate the particle $p_T$ spectra in iHKM.
We analyze the situation at different equations of state for quark-gluon and hadron matter and correspondingly, 
at different temperatures of the so-called chemical freeze-out.

\section{Model description}

The current study is carried out within the 'Integrated Hydrokinetic Model' (iHKM)~\cite{ihkm} of relativistic
nuclear collisions. This model includes the five stages of the matter evolution and observable formation in $A+A$ 
collisions: the initial state formation, the pre-thermal matter evolution, the hydrodynamic stage, the 
particlization and the hadronic cascade stages.
	
The initial energy-density profile in iHKM is associated with a quite early proper time, 
$\tau_0 \approx 0.1$~fm/$c$. According to a combined method, described in~\cite{ihkm}, one presents the generally 
non-equilibrium boost-invariant (in the central region of rapidity) parton/gluon distribution function on the 
initial hypersurface $\sigma_0$: $\tau=\tau_0$ in the following factorized form 
\begin{equation}
f(t_{\sigma_0},\textbf{r}_{\sigma_0},\textbf{p}) = \epsilon(b;\tau_0, {\bf r}_T)f_0(p),
\label{f0}
\end{equation}
where $\epsilon(b;\tau_0, {\bf r}_T)$, being the initial energy density profile, is calculated in a hybrid 
approach, including both wounded nucleon model and the binary collision approach. The proportion between the 
contributions of these two models to $\epsilon(b;\tau_0, {\bf r}_T)$ is regulated by the parameter $0 \leq \alpha \leq 1$. 

In iHKM simulations we obtain the distributions of numbers of wounded nucleons and binary collisions at $\tau_0$ 
with the help of the \textsc{GLISSANDO} code~\cite{Gliss}. 
The weighed sum of such distributions (with the coefficients $\alpha$ and $1-\alpha$) is then multiplied by a 
normalizing factor $\epsilon_0$ --- the energy density at $\tau_0$ in the center of the system in central collisions. 
The value of $\epsilon_0$ is the main free parameter of the model,
defined, together with the parameter $\alpha$, by means of fitting the observed mean charged particle multiplicity 
$dN_{ch}/d\eta$ dependence on centrality at given collision energy. So, both $\epsilon_0$ and $\alpha$ parameters do not depend on collision centrality.
However, changing the equation of state together with the corresponding particlization temperature will require
a modification of $\epsilon_0$ and $\tau_0$ 
parameters~\footnote{As for the parameter $\alpha$, in the current analysis it is not changed when switching to another EoS.}. 
They are fixed in iHKM basing on the measured multiplicity vs centrality distribution and the measured 
slope of the pion transverse momentum spectrum. 
As for the possible momentum anisotropy of parton/gluon initial distribution, typical for the Color-Glass-Condensate-based approaches, it is taken into account by the function $f_0(p)$ in Eq.~(\ref{f0}) which is 
described in more detail in previous papers~\cite{ihkm, ihkm2}:
\begin{eqnarray}
f_0(p)=g \exp\left(-\sqrt{\frac{(p\cdot U)^2-(p\cdot
V)^2}{\lambda_{\perp}^2}+\frac{(p\cdot
V)^2}{\lambda_{\parallel}^2}}\right),
\label{anis1}
\end{eqnarray}
where $U^{\mu}=( \cosh\eta, 0, 0, \sinh\eta)$, $V^{\mu}=(\sinh\eta, 0, 0,\cosh\eta)$. 
In the rest frame of the fluid element one has $\eta=0$, $(p\cdot U)^2-(p\cdot
V)^2=p_{\perp}^2$ and $(p\cdot
V)^2=p_{\parallel}^2$, so that $\lambda_{\parallel}^2$ and $\lambda_{\perp}^2$ can be associated with the two 
effective temperatures --- one along the beam axis and another along the axis, orthogonal to it. 
In such a case the parameter $\Lambda=\lambda_{\perp}/\lambda_{\parallel}$ defines the momentum anisotropy of the initial state. 

Once we have defined the initial conditions in the form of non-thermal energy-momentum tensor, obtained from the 
distribution (\ref{f0}), we can proceed to the description of the pre-thermal matter dynamics, using the 
relaxation time approximation~\cite{ihkm,ihkm1,ihkm2}. This pre-thermal stage starts in iHKM at the initial 
time $\tau_0 \approx 0.1$~fm/$c$, when the initial state is formed, and lasts till the thermalization time 
$\tau_{th}=1$~fm/$c$, when an approximate local thermal equilibrium is supposed to be reached by the initially non-equilibrated system. 

The subsequent matter evolution is described within a relativistic viscous hydrodynamics formalism, with the relativistic current and
energy-momentum tensor in the Israel-Stewart form. We neglect there the bulk viscosity and heat conductivity terms. Since at the LHC energies the baryonic chemical potential in the spatiotemporal region, where the midrapidity observables are formed, is negligibly small, we put it to be just zero.
According to the iHKM results \cite{ihkm} for identified hadron multiplicities, spectra, elliptic flow  and femtoscopy data we put the minimal possible ratio of the shear viscosity coefficient to the entropy density, $\frac{\eta}{s}=\frac{1}{4\pi}$ for the quark-gluon matter.  
The hydrodynamic approximation is justified as long as the matter can be considered remaining close to local 
chemical and thermal equilibrium. But at some temperature $T_p$ both such quasi-equilibrium descriptions get 
destroyed, and the further system's evolution should be described in terms of particles. A switching to such a 
description can be done either gradually or suddenly at $T_p$ isotherm hypersurface. In this paper we utilize the 
latter mode of sharp particlization, comparing the simulation results in the two cases of two different $T_p$ 
values \footnote{In fact, we suppose that the particlization temperature $T_p$ coincides with the temperature when the hadronization process is (almost) completed.}. The construction and treatment of the particlization 
hypersurface in iHKM is realized through the Cornelius routine~\cite{Cornelius}. 

The last stage of system's evolution within iHKM is a hadronic cascade stage, described with the help of
UrQMD model~\cite{urqmd}. At this stage all the particles, previously produced at the particlization stage,
collide and interact with each other, that includes both elastic scatterings and inelastic processes, such as 
baryon-antibaryon annihilation. The unstable particles and resonance states decay (and re-combine) at this 
stage as well.

\section{Results and discussion}

In the current paper we present the results for different particle number ratios and spectra, calculated in iHKM 
at the two different particlization temperatures, $T_{p}=165$~MeV and $T_{p}=156$~MeV, with two corresponding 
equations of state (EoS) for quark-gluon matter --- Laine-Schr\"oder~\cite{EoS} and HotQCD Collaboration --- 
``HotQCD'' EoS~\cite{EoS2}. Using the two equations of state we investigate also whether the form of EoS is 
significant for the description of particle number ratios in the evolutionary model with initial energy density 
$\epsilon(\tau_0)$ as a free parameter. Such a study is important since the extremely high rate of the fireball 
expansion, much larger than in the Early Universe, would lead to modification of effective equation of state as 
compared to the lattice QCD calculations for static system. 
 
\begin{figure}
\centering
\includegraphics[bb=0 0 567 409,width=0.88\textwidth]{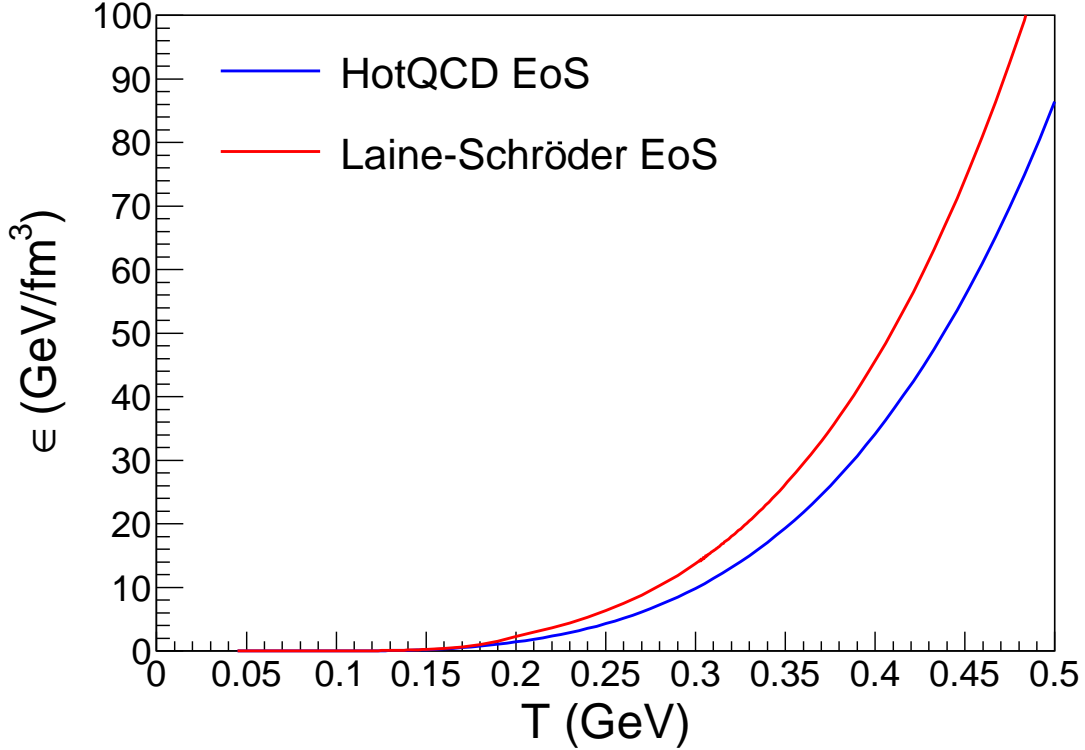}
  \caption{The comparison of two equations of state for quark-gluon matter: the Laine-Schr\"oder EoS~\cite{EoS}, 
  corresponding to the particlization temperature $T_{p}=165$~MeV and the HotQCD Collaboration ``HotQCD'' EoS~\cite{EoS2}, corresponding to the particlization temperature $T_{p}=156$~MeV.
\label{eos}}
\end{figure}

The ratios are calculated for the central ($c=0-10$\%) Pb+Pb collisions at the LHC energy 
$\sqrt{s_{NN}}=2.76$~TeV. The results on particle $p_T$ spectra are demonstrated for the collisions with $c=0-5$\%  
and serve as an additional justification of the choice of model parameters (which do not depend on centrality).
The Laine-Schr\"oder equation of state was previously used in HKM model, the predecessor of iHKM,
as the lattice QCD inspired EoS, ensuring that the description of dense quark-gluon matter and its cross-over type 
transition to the hadron resonance gas pass without gaps in pressure and energy density. 
The corresponding particlization temperature $T_{p}=165$~MeV was used in HKM calculations, that resulted in a
successful simultaneous description of a variety of observables in heavy ion collision experiments at RHIC and LHC 
(spectra, interferometry radii, $v_2$ coefficients, source functions, etc.~\cite{Uniform,Shap,PBM-Sin,Sf,Mtscale}).
The ``HotQCD'' EoS corresponds to the recent HotQCD Collaboration results on lattice QCD simulations
devoted to the quark-gluon matter state description. The respective particlization temperature, $T_p=156$~MeV,
is in agreement with the most recent estimates of the chemical freeze-out temperature
obtained in thermal model, $T_{ch}=156 \pm 1.5$ MeV~\cite{stachel-sqm2013}.
In Fig.~\ref{eos} one can see the comparison of the two EoS on the plot in the coordinates $(\epsilon,T)$.
The Laine-Schr\"oder EoS corresponds to more rapidly growing energy density at the high temperatures.
 
The iHKM parameter values, used in current analysis in the case of $T_p=165$~MeV, are chosen to be the same 
as those that have provided the optimal description of the multiple LHC bulk observables~\cite{ihkm2}: $\tau_0 = 0.1$~fm/$c$, $\tau_{th} = 1$~fm/$c$, the relaxation time at the pre-thermal stage $\tau_{rel} = 0.25$~fm/$c$, $\epsilon_0 = 680$~GeV/fm$^3$, $\alpha=0.24$, the momentum anisotropy of the initial 
state $\Lambda=100$. 
For the new particlization temperature $T_p=156$~MeV most parameter values remain the same, except for $\epsilon_0 = 495$~GeV/fm$^3$ and $\tau_0 = 0.15$~fm/$c$, which are changed in order to ensure the correct charged particle 
multiplicity and pion $p_T$ spectrum slope.

\begin{figure}
\centering
\includegraphics[bb=0 0 567 409,width=0.88\textwidth]{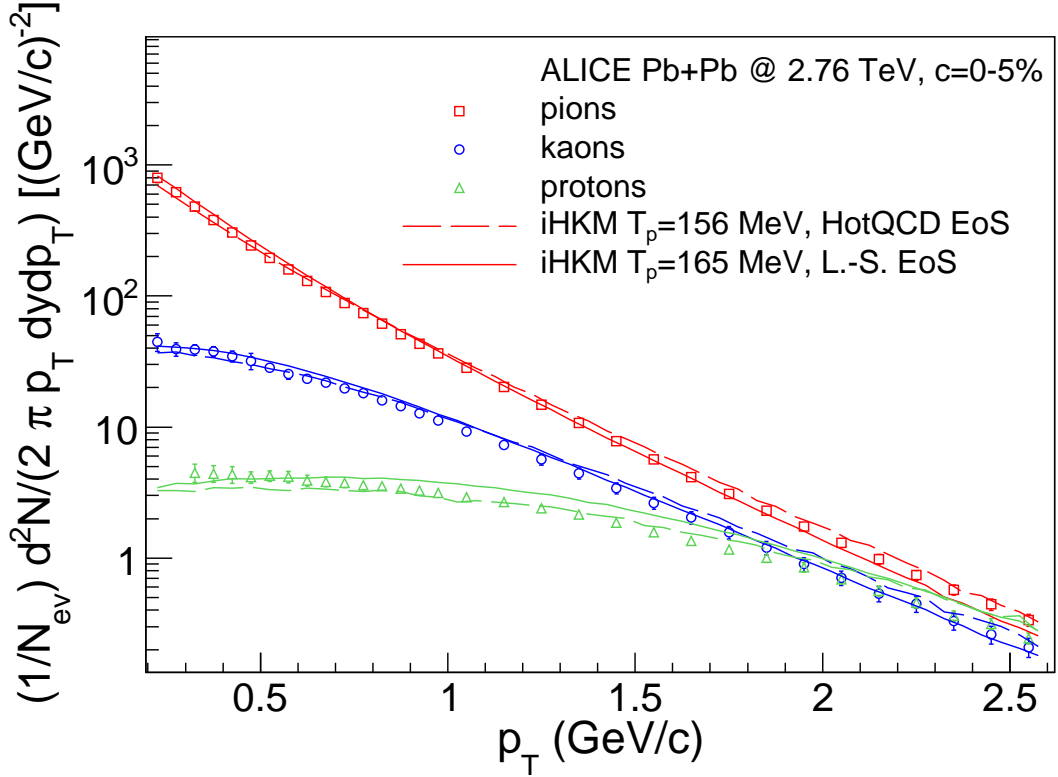}
\caption{The pion, kaon and proton $p_T$ spectra calculated in iHKM model at the two particlization
temperatures, $T_{p}=165$~MeV and $T_{p}=156$~MeV, and corresponding equations of state, \cite{EoS} and~\cite{EoS2} compared with the ALICE experimental data~\cite{ALICE2} for central ($c=0-5\%$) Pb+Pb collisions at the LHC energy $\sqrt{s_{NN}}=2.76$~TeV.
\label{speccmp}}
\end{figure}

In Fig.~\ref{speccmp} one can see the comparison of transverse momentum spectra calculated in iHKM for $c=0-5$\%
Pb+Pb collisions at two mentioned regimes ($T_{p}=165$~MeV and $T_{p}=156$~MeV) together with the experimental 
points. At both particlizaton temperatures the model gives a sufficiently good description of the data, which 
confirms that the model parameters are chosen correctly.

\begin{figure}
\centering
\includegraphics[bb=0 0 567 409,width=0.88\textwidth]{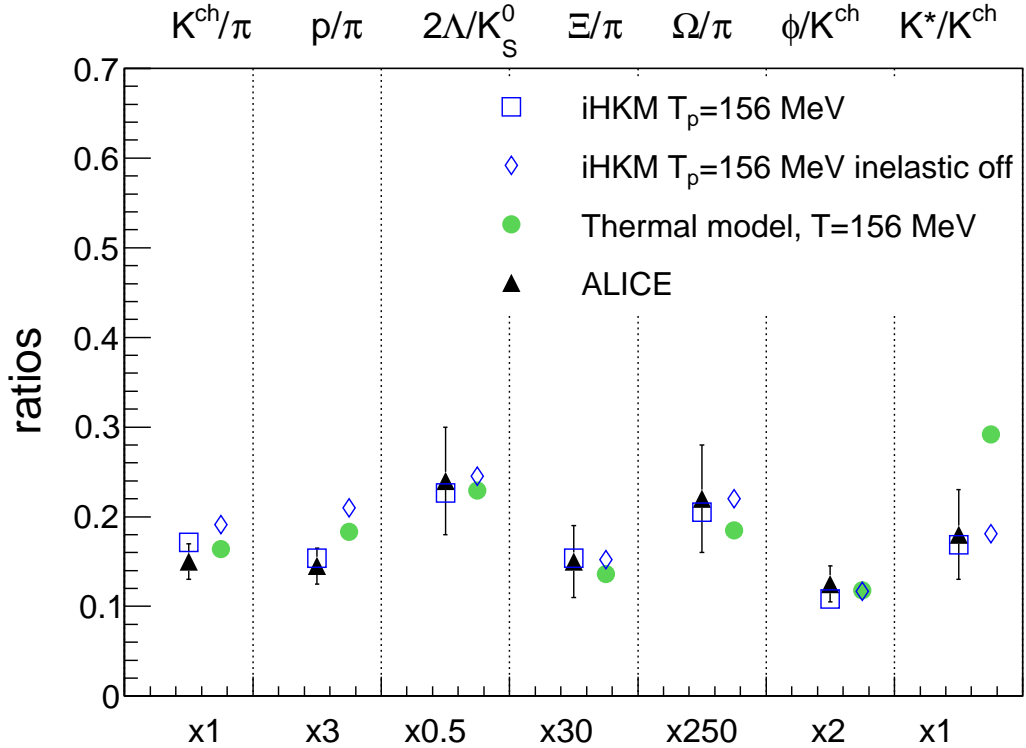}
\caption{The comparison of particle number ratios, calculated in iHKM (blue markers) at the particlization 
temperature $T_{p}=156$~MeV and HotQCD Collaboration  equation of state with the ALICE experimental data~\cite{RatiosExp}
and the thermal model results at $T=156$~MeV~\cite{stachel-sqm2013}. The iHKM simulations are performed in two regimes: full calculation and the mode with inelastic reactions (except for resonance decays) are switched off.
The $\chi^2$ values for these two regimes are 2.2 and 14.9 respectively.
\label{ratioscmp}}
\end{figure}

\begin{figure}
\centering
\includegraphics[bb=0 0 567 409,width=0.88\textwidth]{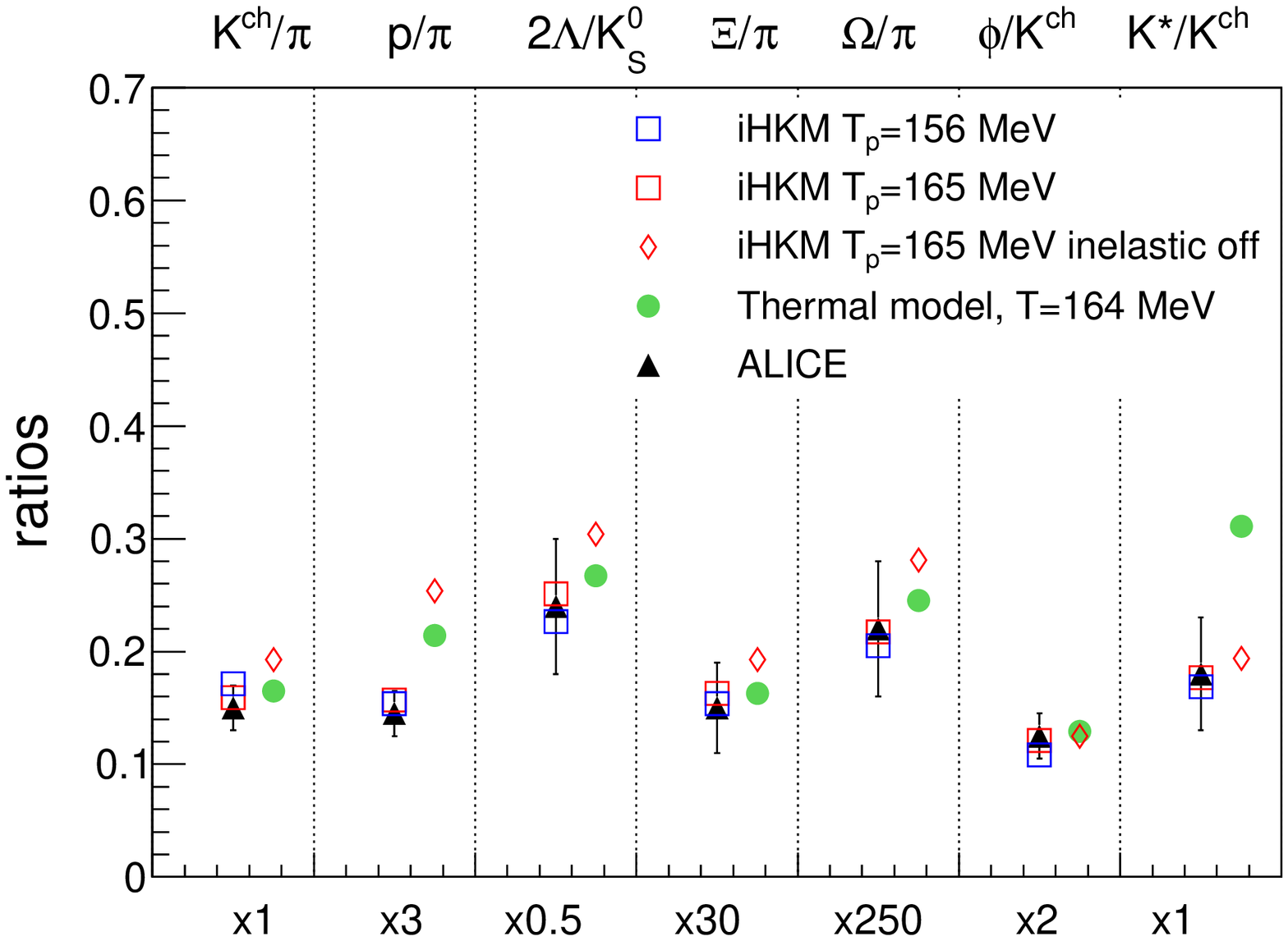}
  \caption{The same as in Fig.~\ref{ratioscmp}, but the results for iHKM calculation at $T_{p}=165$~MeV 
  and Laine-Schr\"oder equation of state are shown.
  The thermal model results are demonstrated for $T=164$~MeV~\cite{stachel-sqm2013, Therm}.
  The $\chi^2$ values for the full and ``switched-off-inelastic'' iHKM simulations are 0.7 and 37.7 respectively. 
\label{ratios5}}
\end{figure}

In Figs.~\ref{ratioscmp}, \ref{ratios5} we demonstrate the iHKM results for a set of particle number ratios and 
compare it with the experimental results and those obtained from the thermal model~\cite{stachel-sqm2013,Therm}.
Here the iHKM simulations are performed in two regimes: full calculation and the mode with the inelastic processes 
switched off (except for resonance decays). It is worth noting that the calculations without inelastic reactions 
but with the initial conditions adjusted to provide right description of the charged hadron multiplicities, 
give the same particle number ratios as without such an adjusting. This effect is clear: when we switch off the 
inelastic reactions (except for the resonance decays), then all the particle numbers on the hypersurface of the 
chemical freeze-out are proportional to the effective volume, $N_i=n_i(T,\mu)V_{eff}$, \cite{Veff}, which 
absorbs the hydro-velocities and space-time characteristics at the chemical freeze-out: $V_{eff}=\int_{\sigma_{ch}} u^{\mu}(x)d\sigma_{\mu}(x)$. 
The same happens with the similarly defined effective volume, related to the unity of rapidity in the case 
of boost invariance in the midrapidity region~\cite{Veff}. Therefore, when one fits the initial energy density 
(and the related initial time) in order to adjust multiplicity distribution at the artificially truncated 
``switched-off-inelastic'' dynamics at the afterburner stage, the only common factor $V_{eff}$ will be modified 
(EoS and corresponding particlization temperature are fixed). So, the particle number ratios will not change,
no matter whether the initial conditions are re-tuned or not. 
  
As one can see from  the figures, the thermal model and the iHKM results, related to the case when the 
inelastic scatterings are switched off~\footnote{Note that some deviation of iHKM results in this truncated case 
from those of the thermal model should be connected with the number of resonances taken into account. In iHKM case 
we consider 329 types of resonances. As for the large deviation in the case of $K^*/K^{ch}$ ratio, it can be explained by different definition of $K^*/K^{ch}$ ratio in the experiment and iHKM from the one side and the thermal model calculations from the other side. As follows from the experimental papers, e. g.~\cite{KstarAlice}, the $K^{*0}(892)$ resonances are reconstructed via the products of their decay into $K^{+}\pi^{-}$ pairs with branching ratio 0.66 (while the $K^{*0}$'s decaying through a channel $K^{*0} \rightarrow K^0\pi^0$ are excluded from the analysis). The same reconstruction procedure is applied in the iHKM study. Hence, the number of 
$K^*$’s, identified in such a way is about 2/3 of the full $K^*$ number. In contrast, the thermal model describes the full $K^*$ number and therefore gives higher $K^*/K^{ch}$ ratio.}, are modified noticeably when the temperature $T_p$ (or $T_{ch}$ in thermal 
models) is changing, and describe the data worse than the full iHKM calculations. As for the latter, they give 
very close results at both particlization temperatures and equations of state!

In a very recent paper~\cite{KstarOur} an essential influence of the particle rescatterings at the 
afterburner stage of the collision on the $K^{*}(892)$ resonance observability was shown. 
It means that the so-called 
thermal freeze-out is not sharp/sudden, but continuous. Our current study points out to the dynamical continuous 
character of the so-called chemical freeze-out in the relativistic heavy ion collisions. It demonstrates that the 
account for inelastic processes at the afterburner stage of the collision plays more important role in the correct  
description of experimental observables, than the specific choice of the supposed particlization/hadronization 
temperature.

\section{Conclusions}

The particle $p_T$ spectra and particle number ratios calculated in iHKM at the two different thermodynamic 
equations of state and corresponding particlization/hadronization temperatures demonstrate that the satisfactory 
description of the experimental data can be achieved at both $T_p$ values if the initial energy density $\epsilon(\tau_0)$ is the free parameter. In this sense the results practically do not depend on the equation of state in complete dynamics of rapidly expanding fireballs formed in A+A collision. However, the situation is different when one truncates the post hydrodynamic stage, the description is better for lower temperature of chemical freeze-out, $T=156$ Mev.  But even in this case --- when annihilation and other inelastic processes (except for the resonance decays) at the afterburner 
stage are neglected --- the theoretical results get worse as compared to the full culculations. One can conclude that 
neither thermal nor chemical freeze-out can be considered as sudden at some corresponding temperatures. 
%Particle number ratios as well as absolute values of particle yields 
Our analysis shows that even at the minimal 
hadronization temperature near 155~MeV, the annihilation and other non-elastic scattering reactions still play 
noticeable role in the formation of particle number ratios, especially those where protons and pions are 
participating.

The fact that the results of iHKM evolutionary model for small and relatively large particlization temperatures 
are quite similar means that inelastic processes (other than the resonance decays), which occur during the 
matter evolution below the corresponding temperature, play a role of the compensatory mechanism in formation of 
the particle number ratios. 

Thus, the current analysis supports the picture of continuous chemical freeze-out at the LHC in the sense that the 
corrections to the sudden chemical freeze-out results, accounting for the inelastic reactions at the  
subsequent times, are important and improve the description of the experimental data.  

%\section*{Acknowledgments}
\begin{acknowledgments}
Yu.S. thanks to P. Braun-Munzinger for fruitful and stimulating discussions. 
The research was carried out within the scope of the EUREA:
European Ultra Relativistic Energies Agreement (European
Research Group: ``Heavy ions at ultrarelativistic energies'', Agreement F-2018 with the National Academy of Sciences (NAS) of Ukraine. 
The work is partially supported by the NAS of Ukraine Targeted research program ``Fundamental research on high-energy physics and nuclear physics (international cooperation)''.
\end{acknowledgments}

\end{document}